\documentclass[journal,]{IEEEtran}
\usepackage[pagewise]{lineno}
\usepackage{cite}
\usepackage{amsmath,amssymb,amsfonts}
\usepackage{graphicx}
\usepackage{textcomp}
\usepackage{xcolor}
\usepackage{algorithm, algorithmic}
\usepackage{subfigure}
\usepackage{booktabs}
\usepackage{multirow}
\usepackage{threeparttable}
\usepackage{makecell}

\def\BibTeX{{\rm B\kern-.05em{\sc i\kern-.025em b}\kern-.08em
T\kern-.1667em\lower.7ex\hbox{E}\kern-.125emX}}
\begin{document}
% \linenumbers
% adding linenumber
%\pagewiselinenumbers
\switchlinenumbers

\title{User-level Social Multimedia Traffic Anomaly Detection with Meta-Learning}

\author{Tongtong~Feng, Qi~Qi, {\it Senior Member, IEEE,} and Jingyu~Wang, {\it Senior Member, IEEE} 
	\thanks{Tongtong Feng is with the department of computer science and technology, Tsinghua University, Beijing 100084, China (e-mail: fengtongtong@tsinghua.edu.cn). Qi Qi and Jingyu Wang are with the State Key Laboratory of Networking and Switching Technology, Beijing University of Posts and Telecommunications, Beijing 100876, China and also with the EBUPT.COM, Beijing 100191, China (e-mail: qiqi8266@bupt.edu.cn; wangjingyu@bupt.edu.cn).}
	}

\markboth{User-level Social Multimedia traffic Anomaly Detection with Meta-Learning}%
{Shell \MakeLowercase{\textit{et al.}}: Bare jrnl of IEEEtran.cls for IEEE Journals}

\maketitle

\begin{abstract}
Accuracy anomaly detection in user-level social multimedia traffic is crucial for privacy security. Compared with existing models that passively detect specific anomaly classes with large labeled training samples, user-level social multimedia traffic contains sizeable new anomaly classes with few labeled samples and has an imbalance, self-similar, and data-hungry nature. Recent advances, such as Generative Adversarial Networks (GAN), solve it by learning a sample generator only from seen class samples to synthesize new samples. However, if we detect many new classes, the number of synthesizing samples would be unfeasibly estimated, and this operation will drastically increase computational complexity and energy consumption. Motivation on these limitations, in this paper, we propose \textit{Meta-UAD}, a Meta-learning scheme for User-level social multimedia traffic Anomaly Detection. This scheme relies on the episodic training paradigm and learns from the collection of K-way-M-shot classification tasks, which can use the pre-trained model to adapt any new class with few samples by going through few iteration steps. Since user-level social multimedia traffic emerges from a complex interaction process of users and social applications, we further develop a feature extractor to improve scheme performance. It extracts statistical features using cumulative importance ranking and time-series features using an LSTM-based AutoEncoder. We evaluate our scheme on two public datasets and the results further demonstrate the superiority of Meta-UAD.
\end{abstract}


\section{INTRODUCTION}
\IEEEPARstart{S}{ocial} multimedia can keep in touch with friends and family, fill spare time, see what's being talked about, find articles and videos, including famous Facebook, YouTube, WhatsApp, WeChat platforms and having become inseparable from people's daily life. Analysis from Kepios\footnote{https://kepios.com/.} shows that there are 4.74 billion social multimedia users around the world in October 2022, equating to 93.4{\%} internet users or 75.4{\%} of the total global population aged 13 and above\footnote{https://datareportal.com/social-media-users/.}. Data from Cisco\footnote{https://www.cisco.com/c/en/us/solutions/collateral/service-provider/visual-networking-index-vni/complete-white-paper-c11-481360.html.} reveals that social multimedia traffic will account for 82{\%} of all Internet traffic by 2022. 

Accuracy anomaly detection in social multimedia traffic is crucial for privacy security\cite{base3, TNSM1, TNSM2, base4, TNSM7}. Conclusion from Report\footnote{https://www.insiderintelligence.com/content/digital-trust-benchmark-report-2021.} counts that 52{\%} social multimedia users are strongly concerned about platforms' protection of their privacy and data. Attackers might exploit applications containing vulnerability, consequently can jeopardize the confidentiality, integrity, and availability of users' crucial information. For example, since users usually register apps according to their interests, their privacy and account information would be compromised if an attacker attacked these vulnerable apps. This might result in an immeasurable financial loss and unrecoverable damage to a person.

\begin{figure*}[t]
	\centering
	\includegraphics[width=0.89\linewidth]{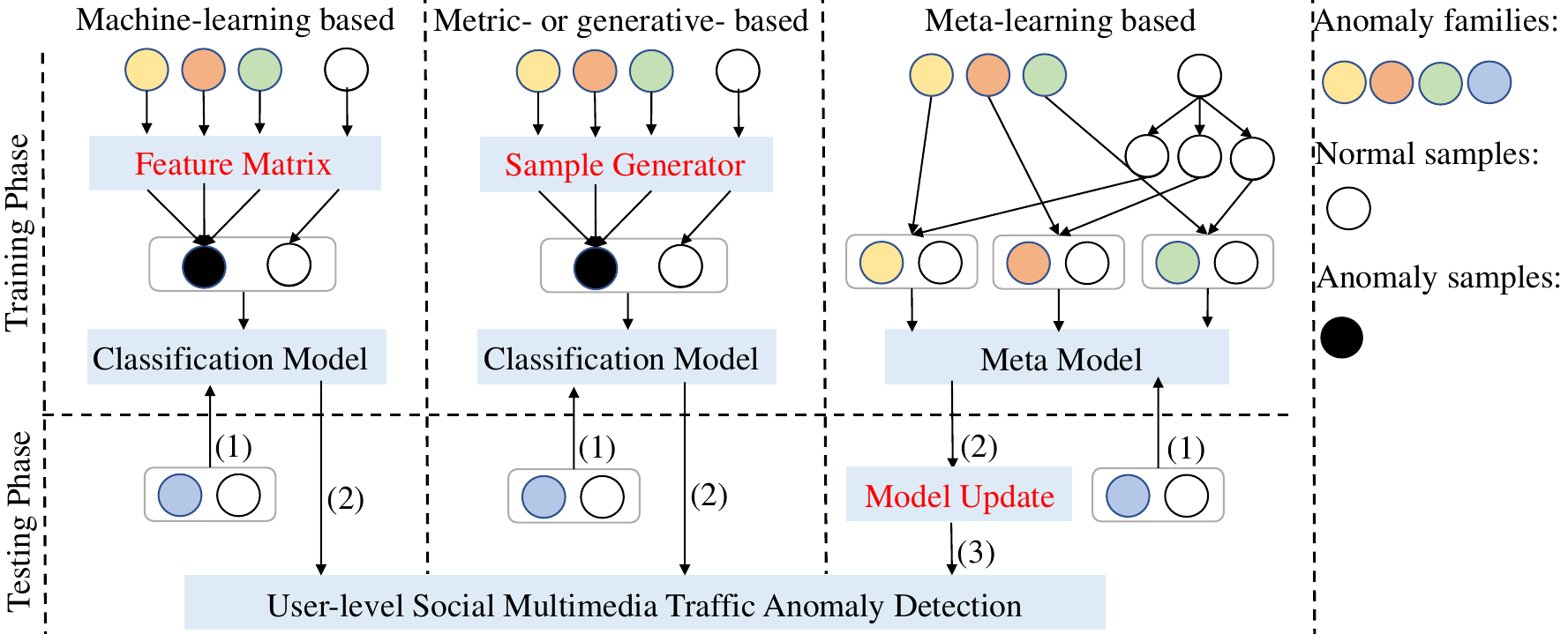}
	\caption{Compared with different social multimedia traffic anomaly detection models.}
	\label{FSL}
\end{figure*}

{\it User-level social multimedia traffic contains sizeable new anomaly classes with few labeled samples}.  As reported by CrowdStrike\footnote{https://go.crowdstrike.com/global-threat-report-2022.html.\label{CrowdStrike}}, about 82{\%} increase in new malware samples related data leaks, 45{\%} increase in the new interactive intrusion campaigns, and adding 21 new anomaly classes in 2021. Those new anomaly classes possess three unique characteristics. {\it Imbalanced}: compare to the existing anomaly classes, it is expensive and arduous to collect a massive amount of data onto the new anomaly class. Therefore, the sample sizes of different anomaly classes in the sampled database are often highly imbalanced. {\it Self-similar}: the new anomaly classes evolve from the existing ones and have characteristics closer to normal traffic. As our detection models adapt, so do anomaly traffics. According to the McAfee labs threats reports\footnote{https://www.mcafee.com/enterprise/en-us/assets/reports/rp-threats-jun-2021.pdf.}, most new anomaly classes are branches of existing anomaly families. {\it Data-hungry}: now more attackers are focusing on fewer but more precise targets instead of widespread invasions, so each new anomaly class has small-scale samples. According to the CrowdStrike report\textsuperscript{\ref{CrowdStrike}}, near 62{\%} attackers will remain silent until a precise target is discovered, and they use legitimate credentials and built-in tools to attack, which can avoid detection by traditional detection models.

Existing anomaly detection models can be grouped into three categories (see in Fig. \ref{FSL}): machine-learning based, metric-learning based, and generative-based models. Machine-learning based models\cite{SVM, DT, RF, CNN, LSTM} classify data using a feature matrix where rows and columns correspond to each packet and feature values. They can combine the rules that differentiate anomaly traffic from normal traffic and achieve high-precision classification. However, they are hard to detect anomaly samples in user-level social multimedia traffic. 1) They heavily rely on large-scale training samples and anomaly class distribution in training datasets and only can achieve good results for specific anomaly classes (same as training dataset). For small-scale training datasets (e.g., endpoint traffic), the rules learned by these models may be entirely inapplicable to other datasets (e.g., cloud traffic). 2) For extremely imbalanced training datasets, models can produce an inductive learning bias towards the majority classes due to overfitting, resulting in poor minority class detection performance. 3) Since the anomaly classes are constantly updated and training data cannot cover all anomaly classes, the effectiveness of existing models will decline sharply for new anomaly classes. 

The metric-learning based models\cite{Simpleshot, TADAM} can learn a non-linear metric to generate new anomaly samples by distance to a class prototype or a class sub-space. Simultaneously, the generative-based models\cite{Delta-encoder, Mfcgan} can learn a sample generator only from seen class samples to synthesize new samples using Generative Adversarial Networks (GAN). These models can greatly alleviate the data-imbalance problem. However, 1) The sample generator learned during their training phase is only for existing anomaly classes. If we detect many new anomaly classes, the number of synthesizing samples would be unfeasibly estimated, and this operation will drastically increase computational complexity and energy consumption. 2) They also do not learn a generic optimal parameter that can easily generalize for seen and unseen class samples. 3) They need a large number of labeled samples for training. 

We argue that the recent breakthroughs in meta-learning have greatly facilitated small-scale sample processing capacities. Among them, few-shot learning\cite{FSL1, FSL2, FSL3, FSL4,base1} enables the model to effectively learn information from small samples by combining limited supervised information (few labels) with prior knowledge (meta-network), which aims to learn representations that generalize well to new tasks. The dominant optimization-based few-shot learning models, such as MAML\cite{MAML}, consist of meta-training and meta-testing phases. The meta-training phase will build a meta-network by training with different sub-tasks. It allows models to learn a common part of different sub-tasks, such as how to extract crucial features. The meta-testing phase will update the pre-trained model to adapt to new tasks by a few iteration steps. Meta-SGD\cite{Meta-SGD} improves MAML with an adaptive learning factor $\alpha$, which can adaptively adjust the update direction and learning rate.

As shown in Fig. \ref{FSL}, compared to existing anomaly detection models, meta-learning does not require generalization to large labeled samples in the training phase and can reduce computational complexity and energy consumption. It can study a meta-model for all anomaly classes and has extremely high generalization. In the testing phase, it can update the meta-model for a specific anomaly class group and has high detection accuracy.

Motivation on this opportunity, in this paper, we propose \textit{Meta-UAD}, a Meta-learning scheme for User-level social multimedia traffic Anomaly Detection, which is based on Meta-SGD\cite{Meta-SGD}, as shown in Fig. \ref{structure}. This scheme relies on the episodic training paradigm and learns from the collection of K-way-M-shot classification tasks, which can use the pre-trained model to adapt new anomaly classes with few labeled samples. Since user-level social multimedia traffic emerges from a complex interaction process of users and social applications, we further develop a feature extractor to improve scheme performance. It extracts statistical features using cumulative importance ranking and time-series features using an LSTM-based AutoEncoder\cite{AutoEncoder1}. To evaluate our scheme's efficiency, we conduct four comparative experiments in two public social multimedia traffic datasets and the results further demonstrate the superiority of Meta-UAD. 

In summary, this paper makes the following key contributions:
\begin{itemize}
	\item We analyze the challenges faced in user-level social multimedia traffic anomaly detection and propose Meta-UAD to solve them. Meta-UAD can efficiently train a meta-model to detect new anomaly classes with few labeled samples.
	\item We develop a feature extractor to improve scheme performance by extracting statistical features using cumulative importance ranking and time-series features using an LSTM-based AutoEncoder.
	\item We implement Meta-UAD and use two public social multimedia traffic datasets for trace-driven experiments. We compare our scheme with the existing models and the results further demonstrate the superiority of Meta-UAD.
\end{itemize}

The rest of this paper is organized as follows. Section \uppercase\expandafter{\romannumeral2} describes the challenge of existing anomaly detection models and the overview of our scheme. Section \uppercase\expandafter{\romannumeral3} presents the design details of Meta-UAD. Section \uppercase\expandafter{\romannumeral4} presents the implementation details of Meta-UAD. Section \uppercase\expandafter{\romannumeral5} presents the evaluation results with a comparison to existing models. Section \uppercase\expandafter{\romannumeral6} briefly reviews the related works. Conclusion and future works are discussed in Section \uppercase\expandafter{\romannumeral7}.

\section{Scheme Overview}

\subsection{Challenge}\label{problem}
We define flow $f = \{p_1, p_2,...,p_u\}$ as a set of adjacent packets $p$ sharing the same 5-tuple information: [source IP address, destination IP address, source port number, destination port number, and transport layer protocol], where $u$ is equal to the number of packet $p$ in flow $f$. We define $\vec{A}_{(p_i)} = {a_j(p_i)}_{0\leq j \leq v}$ as the feature vector of packet $p_i$, where $a_j(p_i)$ is one feature of packet $p_i$, such as the packet size, and $v$ is the feature number of packet $p_i$. We refer to $X^{u,v}_f=\{\vec{A}_{(p_i)}\}_{0\leq i \leq u}$ as the feature matrix of flow $f$. We denote by $y_f$ the detection label of flow $f$. Therefore, each sample in the training dataset can be expressed as ($X^{u,v}_f, y$), and each anomaly detection model can be shown: $\pi_{\theta}: X^{u,v}_f \rightarrow y$.

The existing anomaly detection models can be roughly categorized into machine-learning based and metric- or generator-based models. For machine-learning based models, given a flow $f$, the model $\pi_{\theta}(\cdot)$ can use a classifier to classify it as class $y^{'}$. For metric- or generator-based models, given many samples of $Q$ anomaly classes, those models first learn a sample generator with parameters $\theta$ that takes these samples as its input and balance sample numbers between different classes,  then use a machine-learning based classifier for anomaly detection. The accuracy of each detection model is determined by the entropy between the detection type $y^{'}$ and the actual type $y$.

{\it Existing models hardly adapt to new anomaly classes.} Their goal is to learn a detection model $\pi_{\theta}(\cdot)$ from $Q$ anomaly classes during the training phase and to achieve high-precision detection of the same anomaly classes in the testing phase. The training samples only contain some existing anomaly classes since it is different to collect training samples with all anomaly classes in real-world applications. Sample generators can synthesize new samples from existing class samples, but their most influential role is to balance sample numbers between different classes. If we detect many new classes, the number of synthesizing samples would be unfeasibly estimated, and this operation will drastically increase computational complexity and energy consumption.

{\it Existing models need large labeled training samples.} Machine-learning based classifiers can achieve high-precision classification by combining rules that differentiate anomaly traffic from normal traffic. When only one or few samples per anomaly class are in the training dataset, classifiers will not work or have inferior generalizations. When a few anomaly classes have one or few samples, classifiers will produce an inductive learning bias towards the majority classes due to overfitting, resulting in poor minority class detection performance. Although sample generators can synthesize new samples from seen class samples, sample generators' optimized space is strongly positively correlated with sample distribution for each anomaly class. When the training samples of some anomaly classes are very few, it is challenging to construct a globally optimal and stable sample generator. 

User-level social multimedia traffic contains many new anomaly classes with few labeled samples and has three natures: imbalance, self-similar, and data-hungry. According to the above limitations, existing anomaly detection models are challenging to apply to this scenario. Besides, we also make it explicit that the choice of the detection model $\pi_{\theta}(\cdot)$ depends on feature engineering $X^{u,v}_f$.

\subsection{Scheme Overview}\label{overview}
\begin{figure}[t]
	\centering
	\includegraphics[width=\linewidth]{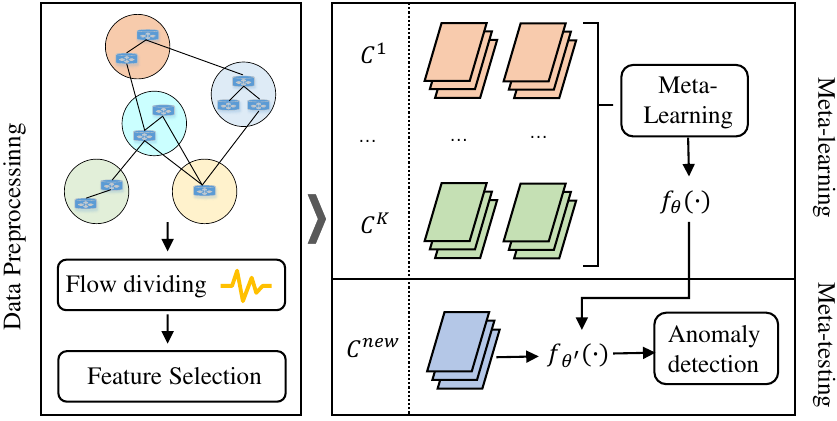}
	\caption{An overview of user-level social multimedia traffic anomaly detection scheme based on Meta-learning.}
	\label{structure}
\end{figure}

To meet this challenge, we propose Meta-UAD (see in Fig. \ref{structure}), which contains two modules, a data preprocessing module and an anomaly detection module. In the data preprocessing module, given some packets as input, Meta-UAD first groups them into flows sharing the same 5-tuple information. Then, it extracts flow features using a feature extractor, including statistical features based on cumulative importance ranking (describe in \ref{Statistical feature}) and time-series features calculated from an LSTM-based AutoEncoder (describe in \ref{Time-series feature}). Finally, it concatenates the 2-part features as the input of the anomaly detection module.

In the anomaly detection module, we design a detector based on Meta-SGD\cite{Meta-SGD} to identify anomaly traffic. Formally, we have a large training dataset $\mathcal{D}^{tr}$ (typically anomaly classes with large samples) and a test dataset $\mathcal{D}^{test}$ (new anomaly classes with one or a few samples), in which their respective class setscolon $\mathcal{C}_{tr}=\{1,...,|\mathcal{C}_{tr}|\}$ and $\mathcal{C}_{test}=\{|\mathcal{C}_{tr}|+1,...,|\mathcal{C}_{tr}|+|\mathcal{C}_{test}|\}$ are disjoint. Our scheme aims to learn a classification model on $\mathcal{C}^{tr}$ that can accurately generalize to new classes $\mathcal{C}_{test}$ with few samples. It consists of a meta-training phase and a meta-testing phase. In each episode of meta-training phase, we first sample $K$ classes from $\mathcal{C}_{tr}$, $\mathcal{C}^K \sim \mathcal{C}_{tr}$. Then, we sample $M$ and $N$ labelled flows per class in $\mathcal{C}^K$ to construct the support set $\mathcal{D}^{sup}=\{(x_m,y_m)_m\}$ and the validation set $\mathcal{D}^{var}=\{(x_n,y_n)_n\}$, respectively. Meanwhile, we construct the task $\mathcal{T}_i = \{\mathcal{D}^{sup}_i,\mathcal{D}^{var}_i\}$ for each class $\mathcal{C}^i$ with a support set $\mathcal{D}^{sup}_i$ and a validation set $\mathcal{D}^{var}_i$. $\mathcal{D}^{sup}_i$ is used for inner update through gradient descent and a learning factor $\alpha$ to obtain the updated parameters $\theta'_i$ for each task. The learning factor $\alpha$ can adaptively adjust the update direction and learning rate. Then $\mathcal{D}^{var}_i$ is used to measure the performance of $\theta'_i$. An outer update procedure is used to update the model parameters $\theta$ and the learning factor $\alpha$ by taking into account all the sampled tasks. In the meta-testing phase, given a new class $\mathcal{C}^{new} \sim \mathcal{C}_{test}$, we use only a few flows to get the adapted parameters $\theta'$ for this specific anomaly class $\mathcal{C}^{new}$. In the process of model construction, specific details such as task definition, meta-training, meta-testing, and backbone architecture, are shown in Section \ref{Meta-SGD algorithm}.

\section{Scheme Design}
\subsection{Statistical features extraction}\label{Statistical feature}
In this subsection, we first use CICFlowMeter\cite{CICFlowMeter} to extract 81 flow-level statistical features and define them as {\it feature{\_}set{\_}1}, as shown in TABLE \ref{feature set 1}. 

\begin{table}[t]
	\centering
	\caption{feature{\_}set{\_}1}
	\label{feature set 1}
	\begin{threeparttable}
		\setlength{\tabcolsep}{2.5mm}{
		\begin{tabular}{|c|l|}
			\hline
			No. & Feature Name  \\ \hline
			1 & Flow duration \\ \hline
			2 & Flow byte/s: Mean  \\ \hline
			3 & Fwd packets/s: Mean \\ \hline
			4 & Bwd packets/s: Mean \\ \hline
			5 & Flow packets/s: Mean \\ \hline
			6-10 & Fwd packet count: Total, Min, Max, Mean, Std \\ \hline
			11-15 & Bwd packet count: Total, Min, Max, Mean, Std  \\ \hline
			16-20  & Flow packet count: Total, Min, Max, Mean, Std \\ \hline
			21 & Bwd/Fwd packet total count ratio \\ \hline
			22-26 & Fwd header Length: Total, Min, Max, Mean, Std \\ \hline
			27-31 & Bwd header Length: Total, Min, Max, Mean, Std \\ \hline
			32-36 & Flow header Length: Total, Min, Max, Mean, Std \\ \hline
			37 & Bwd/Fwd header total length ratio \\ \hline
			38-42 & Fwd packet Length: Total, Min, Max, Mean, Std  \\ \hline
			43-47 & Bwd packet Length: Total, Min, Max, Mean, Std \\ \hline
			48-52 & Flow packet Length: Total, Min, Max, Mean, Std \\ \hline
			53 & Bwd/Fwd packet total length ratio \\ \hline
			54-58 & Fwd IAT: Total, Min, Max, Mean, Std \\ \hline
			59-63 & Bwd IAT: Total, Min, Max, Mean, Std \\ \hline
			64-68 & Flow IAT: Total, Min, Max, Mean, Std\\ \hline
			69 & Bwd/Fwd total IAT ratio \\ \hline
			70-71 & Fwd flag count: PSH, URG \\ \hline
			72-73 & Bwd flag count: PSH, URG  \\ \hline
			74-77 & Flow flag count: FIN, SYN, RST, PSH \\ \hline
			78-81 & Flow flag count: ACK, URG, CWR, ECE \\ \hline
		\end{tabular}}
	\begin{tablenotes}
		\footnotesize
		\item[1] Fwd/Bwd: forward/backward flow; 
		\item[2] IAT: the sending time interval between two adjacent packets. 
		\item[3] Flag count: number of packets with this flag.
	\end{tablenotes}
	\end{threeparttable}
	
\end{table}

\begin{table}[t]
	\centering
	\caption{feature{\_}set{\_}2}
	\label{feature set 2}
		\begin{tabular}{|c|l|}
			\hline
			No. & Feature Name  \\ \hline
			1-2 & Bwd packet count: Total, Mean  \\ \hline
			3-4  & Flow packet count: Total, Mean \\ \hline
			5 & Bwd/Fwd packet total count ratio \\ \hline
			6 & Fwd header Length: Total \\ \hline
			7 & Bwd header Length: Total \\ \hline
			8 & Flow header Length: Total \\ \hline
			9 & Bwd/Fwd header total length ratio \\ \hline
			10-13 & Fwd packet Length: Total, Max, Mean, Std  \\ \hline
			14-17 & Bwd packet Length: Total, Max, Mean, Std \\ \hline
			18-21 & Flow packet Length: Total, Max, Mean, Std \\ \hline
			22 & Bwd/Fwd packet total length ratio \\ \hline
			23-26 & Fwd IAT: Min, Max, Mean, Std \\ \hline
			27-28 & Bwd IAT: Max, Mean \\ \hline
			29-31 & Flow IAT: Total, Max, Mean\\ \hline
			32 & Fwd flag count: PSH \\ \hline
			33 & Flow flag count: ACK \\ \hline
	\end{tabular}
\end{table}

By experiment testing, we find that some features are equal to $0$, and some features are missing in some flows. So we use three strategies for feature selection: the proportion of missing values, the entropy, and the cumulative importance. We assume $a_j$ is a feature from feature{\_}set{\_}1. When the feature's missing proportion is greater than 50{\%} or the feature's entropy is equal to 0, the feature $a_j$ is deleted. Because different classification algorithms have a large difference in the order of feature importance, we use the comprehensive cumulative importance ranking of five different algorithms. First, the accuracy $Acc_{i}$ of each algorithm $i$ is obtained by running dataset separately in five different algorithms, including RF, Extraboost, Admboost, GBDT, and Lightgbm algorithm. Then the feature importance value of each feature $V_{(a_j)}^i$ is calculated in five different algorithms. Finally, the cumulative feature importance of each feature $C_{(a_j)}$ is calculated as follows: 
\begin{equation}
C_{(a_j)}=\frac{1}{5}\sum_{i=1}^{5}Acc_{i}V_{(a_j)}^i
\end{equation}
where $i$ represents the different algorithms, $i\in\{1,2,..,5\}$ and $a_j$ represents different features, $j \in \{1,2,...,81\}$. Finally, through experimental tests, anomaly detection accuracy reaches the highest when the last 30{\%} features are deleted in features' importance sorting. Through three strategies, 33 features were selected as the basic statistical features ({\it feature{\_}set{\_}2}). These features are shown in TABLE \ref{feature set 2}.

\subsection{Time-series features extraction}\label{Time-series feature}
The time series features can characterize the interaction process of different tasks well. In this subsection, we introduce how to use an LSTM-based AutoEncoder\cite{AutoEncoder1} to extract time-series features in social multimedia traffic.

\textit{Feature matrix.} We first construct a flow feature matrix $X^{u,v}_f=\{\vec{A}_{(p_i)}\}_{0\leq i \leq u}$ (describe in \ref{problem}) as the input of the AutoEncoder. Among them, each flow $f$ contains $u$ packets and each packet $p_i$ contains $V$ packet-level features $\vec{A}_{(p_i)}$: [header length, payload length, packet interval, window size, ack{\_}cnt, pst{\_}cnt, direction]. We do not select 5-tuple values as the packet feature because the IP address and port number will change frequently. Since the number of packets between different flows is not equal, we select $B$ as the maximum number of packets in the feature matrix $X^{u,v}_f$. If the number of packets in the flow $i$ exceeds $B$, we will select the first $N$ packets to construct the feature matrix. In contrast, we will fill the feature matrix by increasing the $\vec{\mathbf{0}}$ vector.

\begin{figure}[t]
	\centering
	\includegraphics[width=0.95\linewidth]{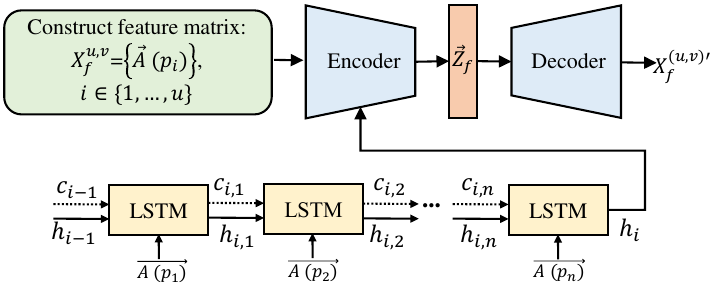}
	\caption{Time-series features extraction based on an LSTM-based AutoEncoder.}
	\label{LSTM}
\end{figure}

\textit{LSTM-based AutoEncoder.} 
We use an LSTM-based AutoEncoder to extract the time-series features of each flow. The design details are as follows. Both the encoder and the decoder use an LSTM network. The input of the encoder is the flow feature matrix $X^{u,v}_f$, and the output is a $z$-dimensional vector $\vec{Z}_f$($|\vec{Z}_f|=z$). The input of the decoder is a vector $\vec{Z}_f$, and the output is a matrix $X^{(u,v)'}_f$, which has the same matrix structure as the encoder input. The AutoEncoder's iterative mechanism is to reduce the mean square error between the input of the encoder and the output of the decoder:
\begin{equation}
J_{(W,b)}=\frac{1}{F}\sum_{u=1}^{U}\sum_{v=1}^{V}(X^{u,v}_f-X^{(u,v)'}_f)^2
\end{equation}
where $W$ and $b$ are the network hyperparameter; $X^{u,v}_f$ represents the $v$-th feature of the $u$-th packet in the flow $f$; $F$ represents the number of flows we consider; We set {\it feature{\_}set{\_}3} as the collection both of feature{\_}set{\_}2 and $z$-dimensional time-series features.

\subsection{Meta-Learning for Anomaly Detection}\label{Meta-SGD algorithm}
MAML\cite{MAML} is the most famous optimization-based meta-learning algorithm, which can study a meta-model from large labeled training samples and adapt it to a new class with few samples. Its optimizer mainly includes initial parameters, update direction, and learning rate. Among them, its initial parameters are often set randomly, the update direction is often the direction of the natural gradient, and the learning rate is often low. When the update direction is a practical direction for data fitting, it can not stop the learning process to avoid overfitting. Meta-SGD\cite{Meta-SGD} has solved these problems well by using a variable learning factor $\alpha$, which can adaptively adjust the update direction and learning rate. Therefore, we use the Meta-SGD structure (describe in Section \ref{overview}) to train a model from only a few anomaly samples to adapt new anomaly classes in user-level social multimedia traffic anomaly detection.

\begin{figure}[b]
	\centering
	\includegraphics[width=0.99\linewidth]{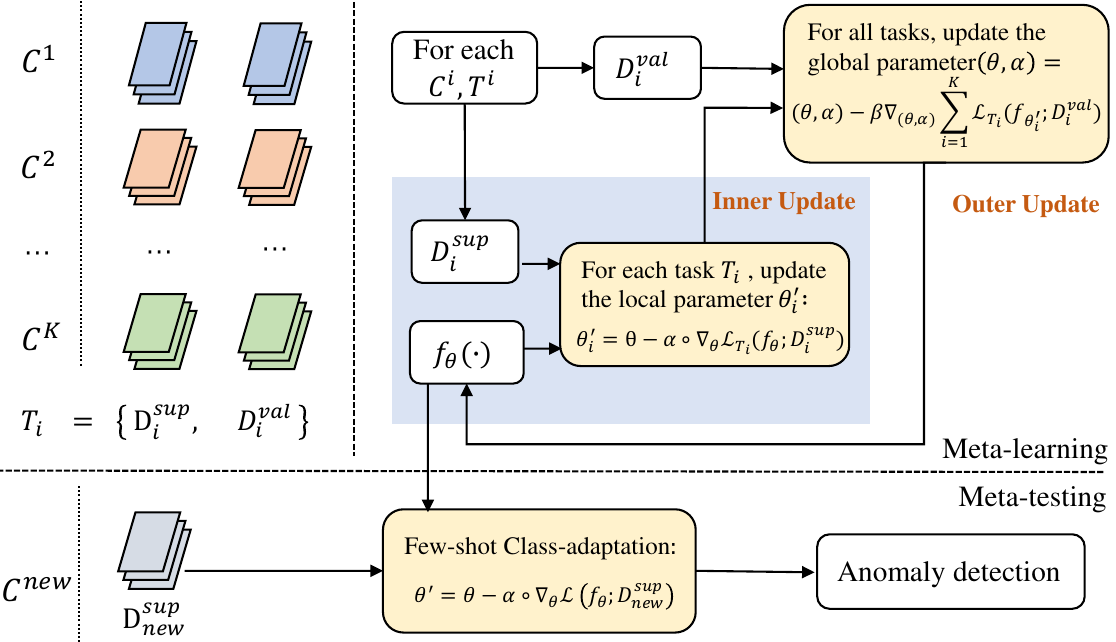}
	\caption{The design details of user-level social multimedia traffic anomaly detection based on meta-learning.}
	\label{meta-learning}
\end{figure}

{\bf Task Definition.} The first challenge is how to define a task, which is usually generated by randomly selecting samples of $K$ classes in each episode. In our scheme, for a given anomaly class $\mathcal{C}^i$, we definite a task as $\mathcal{T}_i = \{\mathcal{D}^{sup}_i,\mathcal{D}^{val}_i\}$, where $\mathcal{D}^{sup}_i$ and $\mathcal{D}^{val}_i$ are the support set and validation set in the task $\mathcal{T}_i$. Among them, $\mathcal{D}^{sup}_i$ and $\mathcal{D}^{val}_i$ contain one specific anomaly class $\mathcal{C}^i$ samples and normal samples. The goal of each task is to identify the specific anomaly class $\mathcal{C}^i$. Suppose we build $K$ tasks in the training phase (see in Fig. \ref{meta-learning}).

{\bf Meta-learning.} The goal of meta-learning is to construct a pre-trained anomaly detection model $\pi_{(\theta,\alpha)}:x \rightarrow y$ with parameters $\theta,\alpha$. $\theta$ is a set of the model's initial parameter; $\alpha$ is a vector that decides both the update direction and learning rate.  Meta-learning phase includes two modules: the inner update and the outer update. Algorithm \ref{algorithm 1} shows the process of meta-learning by the pseudocode.

{\it In the inner update.} The goal of the inner update is to update the local parameters $\theta'_i$ for each task $\mathcal{T}_i$ through gradient descent and the learning factor $\alpha$. For a given anomaly class $\mathcal{C}^i$ and task $\mathcal{T}_i$, we first definite a loss function on the support set $\mathcal{D}^{sup}_i$:
\begin{equation}
\mathcal{L}_{\mathcal{T}_i}(\pi_{\theta};\mathcal{D}^{sup}_i) = \sum_{(x_i,y_i)\in \mathcal{D}^{sup}_i}L(\pi_{\theta}(x_i),y_i)
\end{equation}
where $\pi_{\theta}(x_i)$ is a probability vector that shows the probability of sample $x_i$ being detected as different classes; $y_i$ is a one-hot vector that shows the actual class of the sample $x_i$; $|\pi_{\theta}(x_i)|=|y_i|=K+2$, it means that there are $K+2$ detection results. Among them, there are $K$ anomaly classes in training tasks, 1 normal class, and 1 anomaly class in testing task. We define $L(\cdot)$ is the cross entropy error function:
\begin{equation}
L(\pi_{\theta}(x_i),y_i) = -\sum_{k=1}^{K+2}y_i^klog(\pi_{\theta}(x_i)^k)
\end{equation}
where $\pi_{\theta}(x_i)^k$ is the probability that sample $x_i$ is predicted to the class $k$. Then, we use one gradient update and the vector $\alpha$ to change the parameters from $\theta$ to $\theta'_i$: 
\begin{equation}
\theta'_i=\theta - \alpha \circ \bigtriangledown_{\theta}\mathcal{L}_{\mathcal{T}_i}(\pi_{\theta};\mathcal{D}^{sup}_i)
\label{E1}
\end{equation}
where $\alpha$ is a vector, the direction of $\alpha \circ \bigtriangledown_{\theta}\mathcal{L}_{\mathcal{T}_i}(\pi_{\theta};\mathcal{D}^{sup}_i)$ represents the direction of gradient update, and the length of $\alpha \circ \bigtriangledown_{\theta}\mathcal{L}_{\mathcal{T}_i}(\pi_{\theta};\mathcal{D}^{sup}_i)$ represents the learning rate. In the inner update, the updated parameters $\theta'$ are specifically adapted to the task $\mathcal{T}_i$.

{\it In the outer update.} The goal of the outer update is to update the global parameters $\theta,\alpha$ by taking into account of all the sampled tasks. For a given anomaly class $\mathcal{C}^i$ and task $\mathcal{T}_i$, we also definite a loss function on the validation set $\mathcal{D}^{val}_i$:
\begin{equation}
\mathcal{L}_{\mathcal{T}_i}(\pi_{\theta'_i};\mathcal{D}^{val}_i) = \sum_{(x_i,y_i)\in \mathcal{D}^{val}_i}L(\pi_{\theta'_i}(x_i),y_i)
\label{E2}
\end{equation}

Then, we use one gradient update to change the parameters from $\theta'_i$ to $\theta$ by taking into account of all the sampled tasks:
\begin{equation}
{(\theta, \alpha)} ={(\theta, \alpha)}-\beta\sum_{i=1}^K\bigtriangledown_{(\theta, \alpha)}\mathcal{L}_{\mathcal{T}_i}(\pi_{\theta'_i};\mathcal{D}^{val}_i)
\end{equation}
where $\beta$ is the learning rate of the outer update.

{\bf Meta-testing.} After meta-training, we obtain the pre-trained model parameters $\theta$. During meta-testing, for a given new target anomaly class $\mathcal{C}^{new}$, we first obtain the adapted parameters $\theta'$ with a few fine-tuning steps by the gradient update:
\begin{equation}
{\theta'} ={\theta}-\alpha \circ \bigtriangledown_{\theta}\mathcal{L}(\pi_{\theta}(\cdot))
\end{equation}

Then, we use this model $\pi_{\theta'}(\cdot)$ for anomaly detection in new class $\mathcal{C}^{new}$. 

\begin{algorithm}[t]
	\caption{User-level social multimedia traffic anomaly detection scheme with Meta-SGD}
	\label{algorithm 1}
	{\bf Input:} Hyper-parameters $\beta$; \\
	{\bf Output:} Hyper-parameters $\theta$, $\alpha$;
	\begin{algorithmic}[1]
		\STATE Randomly initialize parameter $\theta, \alpha$ with a meta model $\pi_{(\theta,\alpha)}(\cdot)$;
		\WHILE {not done}
		\STATE Sample $K$ classes from $\mathcal{C}_{tr}$, $\mathcal{C}^K=\{\mathcal{C}^i\}_{i=1}^K$, $\mathcal{C}^K \sim \mathcal{C}_{tr}$;
		\FOR{each $C^i$}
		\STATE Construct the support set $\mathcal{D}^{sup}$ by sampling $M$ labelled flows from class $\mathcal{C}^i$;
		\STATE Construct the validation set $\mathcal{D}^{val}$ by sampling $N$ labelled flows from class $\mathcal{C}^i$;
		\STATE Construct the task $\mathcal{T}_i = \{\mathcal{D}^{sup}_i,\mathcal{D}^{val}_i\}$ for class $\mathcal{C}^i$;
		\STATE Compute the loss function $\mathcal{L}_{\mathcal{T}_i}(\pi_{\theta};\mathcal{D}^{sup}_i)$;
		\STATE Update the local parameter:\\ $\theta'_i=\theta - \alpha \circ \bigtriangledown_{\theta}\mathcal{L}_{\mathcal{T}_i}(\pi_{\theta};\mathcal{D}^{sup}_i)$;
		
		\ENDFOR
		\STATE Update the global parameter by taking into account of all the sampled tasks:\\ ${(\theta, \alpha)} ={(\theta, \alpha)}-\beta\sum_{i=1}^K\bigtriangledown_{(\theta, \alpha)}\mathcal{L}_{\mathcal{T}_i}(\pi_{\theta'_i};\mathcal{D}^{val}_i)$;
		\ENDWHILE 
	\end{algorithmic}
\end{algorithm}

\begin{table}[b]
	\begin{minipage}[!t]{\columnwidth}
		\renewcommand{\arraystretch}{1.3}
		\caption{Performance comparison of the exponential function with $B$.}
		\label{N}
		\setlength{\tabcolsep}{3.8mm}{
			\begin{tabular}{cccccc}
				\toprule
				B & 5 & 10 & 20 & 30 & 40\\
				\midrule
				Accuracy&0.913&0.969&0.991&0.985&0.982\\
				\bottomrule
		\end{tabular}}
	\end{minipage}
\end{table}

{\bf Backbone architecture.} Our user-level social multimedia traffic anomaly detection model is general. In theory, we can use any anomaly detection network as the backbone architecture. Because the data format of social multimedia traffic is irregular, the features of social multimedia traffic have been extracted preliminarily through the data preprocessing module. For the anomaly detection module, we only need a simple backbone architecture to learn good accuracy by using our scheme. So we come up with the Deep Neural Networks (DNN\cite{DNN}) as backbone architecture.

\section{Implementation}
In this section, we delve in-depth into implementing Meta-UAD's every component and describe the best hyper-parameters of the LSTM-based AutoEncoder and the meta-learning based anomaly detection model.

{\it The LSTM-based AutoEncoder.} First, we set that $\vec{A}_{(p_i)}$ contains 8 packet-level features ($V=8$): [packet header length, payload length, packet interval, window size, ack{\_}cnt, pst{\_}cnt, direction]. We do not select 5-tuple values as the packet feature because the IP address and port number will change frequently. We use sufficient training samples to test the packet numbers' impact on the detection model, as shown in TABLE \ref{N}. So we select $B=20$ as the maximum packet number in the feature matrix $X^{u,v}_f$. When $B<20$, many important data packets will be missing, making it difficult to detect anomalies accurately. When $B>20$, there will be many $\vec{0}$ in the feature matrix, making our model overfitting. The network structure of the encoder contains two LSTM layers. The number of hidden cells in the first and second layers is 256 and 128, respectively. Results from these layers are then aggregated in a hidden layer that uses five neurons to apply the sigmoid function. The network structure of the decoder also contains two LSTM layers. The first and second layers include 128 and 256 hidden cells, respectively. However, its final output is a matrix. We use mean square error as a loss function for model optimization in training. Finally, we will use a 5-dimensional vector ($|\vec{Z}_f|=5$) to represent the time-series features for each flow.

\begin{table*}[t]
	\caption{Statistics of Attacks in CIC-AndMal2017.}
	\centering
	\label{CIC-AndMal2017_1}
	
		\begin{tabular}{cc|cc|cc|ccc}
			\toprule
			Scareware       & Flow\_num & Ransomware  & Flow\_num & Adware   & Flow\_num & SMSmalware & Flow\_num  \\
			\midrule
			FakeTaoBao      & 33299     & Koler       & 44555     & Youmi    & 36035     & FakeMart   & 6401       \\
			FakeAV          & 40089     & Pletor      & \bf{4715}      & Selfmite & 13029     & Zsone      & 9644       \\
			AVpass          & 40776     & RansomBO    & 39859     & Shuanet  & 39271     & Beanbot    & 16088      \\
			AndroidDefender & 56440     & Jisut       & 25672     & Gooligan & \bf{93772}     & Nandrobox  & 44517      \\
			VirusShield     & 27994     & PornDroid   & 46082     & Kemoge   & 38771     & Plankton   & 39765      \\
			FakeAppAL       & 44563     & Charger     & 39551     & Koodous  & 32547     & Fakeinst   & 15026      \\
			AVforAndroid    & 42448     & Simplocker  & 36340     & Ewind    & 43374     & SMSsniffer & 33618      \\
			Penetho         & 24459     & Lockerpin   & 25307     & Feiwo    & 56632     & FakeNotify & 25918      \\
			AndroidSpy      & 25414     & WannaLocker & 32701     & Dowgin   & 39682     & Biige      & 34098      \\
			FakeApp         & 34676     & SVpeng      & 54161     & Mobidash & 31034     & Jifake     & 5993       \\
			FakeJobOffer    & 30683     & -            & -          &-          & -          & Mazarbot   & 6065      \\
			\bottomrule
	\end{tabular}
\end{table*}

{\it Backbone architecture.} We use DNN as the backbone architecture for our scheme, which contains three fully-connected layers. The number of hidden cells in the first layer is $256$, and the number in other layers is $128$. Outputs from these layers are then aggregated in a hidden layer that uses $K+2$ (equal to the number of action) neurons to apply the loss function. We fix the hyperparameters $\beta$ in meta-learning at $0.001$. During meta-training, we sample the batch size $K$ of classes in each epoch to be 5; we sample the batch size $M$ and $N$ of flows in each inner update as the same (i.e., $M=N$).

\section{Evaluation}
This section compares our scheme against existing models covering two public social multimedia traffic datasets.

\subsection{Methodology} \label{method}
{\it Datasets.} Experiments are conducted based on two public datasets: CIC-AndMal2017 and CIC-IDS2017.

\begin{figure}[t]
	\centering
	\includegraphics[width=0.99\linewidth]{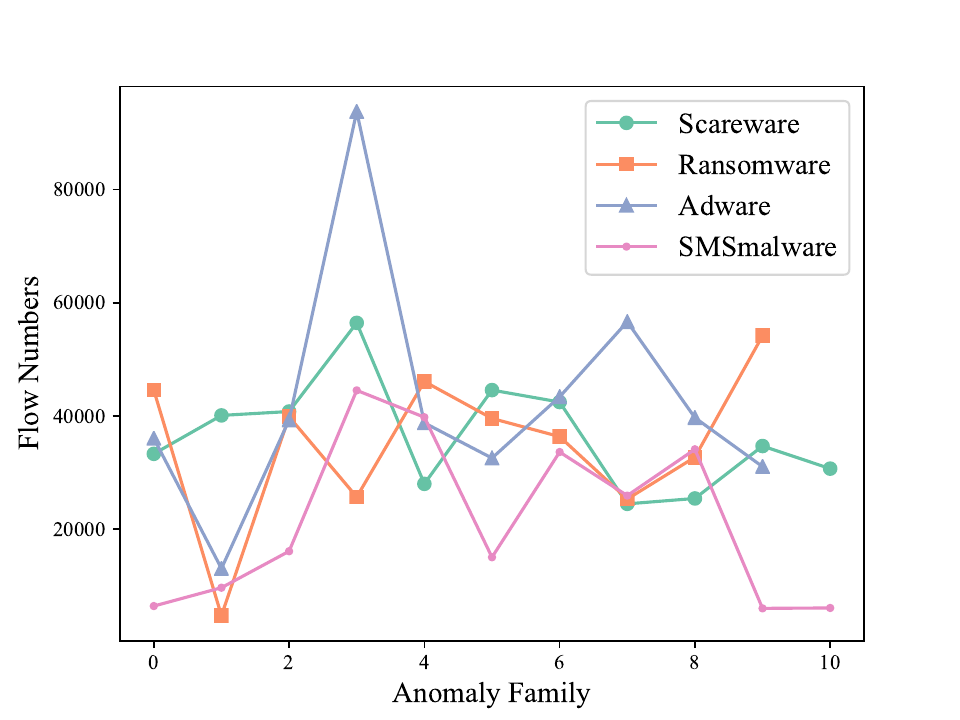}
	\caption{Statistics of Attacks in CIC-AndMal2017.}
	\label{CIC-AndMal2017_2}
\end{figure}

CIC-AndMal2017\cite{CICAndMal2017} is gathered through executing 5065 benign and 429 malware Apps on an actual smartphone instead of an emulator. The benign APPs are collected from the Google Play market published in 2015-2017. The malware Apps are collected from 4 main malware categories (see in TABLE \ref{CIC-AndMal2017_1}): scareware (11 malware families), ransomware (10 malware families), adware (10 malware families), and SMSmalware (11 malware families). On average, each malware family contains ten malware apps. During the data collection process, anomaly traffic was collected in units of malware families instead of each app. Finally, 1205515 normal flows and 1411064 anomaly flows were captured. This dataset can be viewed as imbalanced from Fig. \ref{CIC-AndMal2017_2}. The imbalanced ratios of normal flows to Pletor, FakeJobOffer, AndroidDefender, and Gooligan are 1:255, 1:39, 1:21, and 1:13, respectively.

\begin{table}[t]
	\caption{Statistics of Attacks in CIC-IDS2017.}
	\centering
	\label{CIC-IDS2017}
	\begin{tabular}{ccc}
		\toprule
		Class         & Sample Number & Imbalanced Ratio  \\
		\midrule
		Normal        & 2271320       & -                 \\
		DoS           & 251712        & 1:9               \\
		PortScan      & 158804        & 1:14              \\
		DDoS          & 128025        & 1:17              \\
		FTP-Patator   & 7935          & 1:286             \\
		SSH-Patator   & 5897          & 1:385             \\
		Botnet        & 1956          & 1:1161            \\
		Brute Fore    & 1507          & 1:1507            \\
		XSS           & 652           & 1:3483            \\
		Infiltration  & 36            & 1:63092           \\
		SQL Injection & 21            & 1:108158          \\
		HeartBleed    & 11            & 1:206483          \\
		\bottomrule
	\end{tabular}
\end{table}

CIC-IDS2017\cite{CICIDS2017} is provided by the Canadian Institute of Cybersecurity, which is the latest labeled intrusion detection dataset based on social multimedia platforms and cover the necessary conditions for updated attacks in DoS, PortScan, DDoS, FTP-Patator,SSH-Patator, Botnet, Brute Force, XSS, Infiltration, SQL Injection, HeartBleed (see in TABLE \ref{CIC-IDS2017}). The data capturing period started at 9 a.m., Monday, July 3, 2017, and ended at 5 p.m. on Friday, July 7, 2017, for 5 days. Finally, 2271320 normal flows and 556556 anomaly flows were captured. Compared to CIC-AndMal2017, it is an extremely imbalanced dataset and it contains many novel attacks. From the statistical details in TABLE \ref{CIC-IDS2017}, the imbalanced ratios of normal flows to HeartBleed, XSS, and DoS are 1:206483, 1:3483 and 1:9, respectively.

\begin{figure*}[t]
	\centering 
	\subfigure[]{
		\centering
		\includegraphics[width=0.32\linewidth]{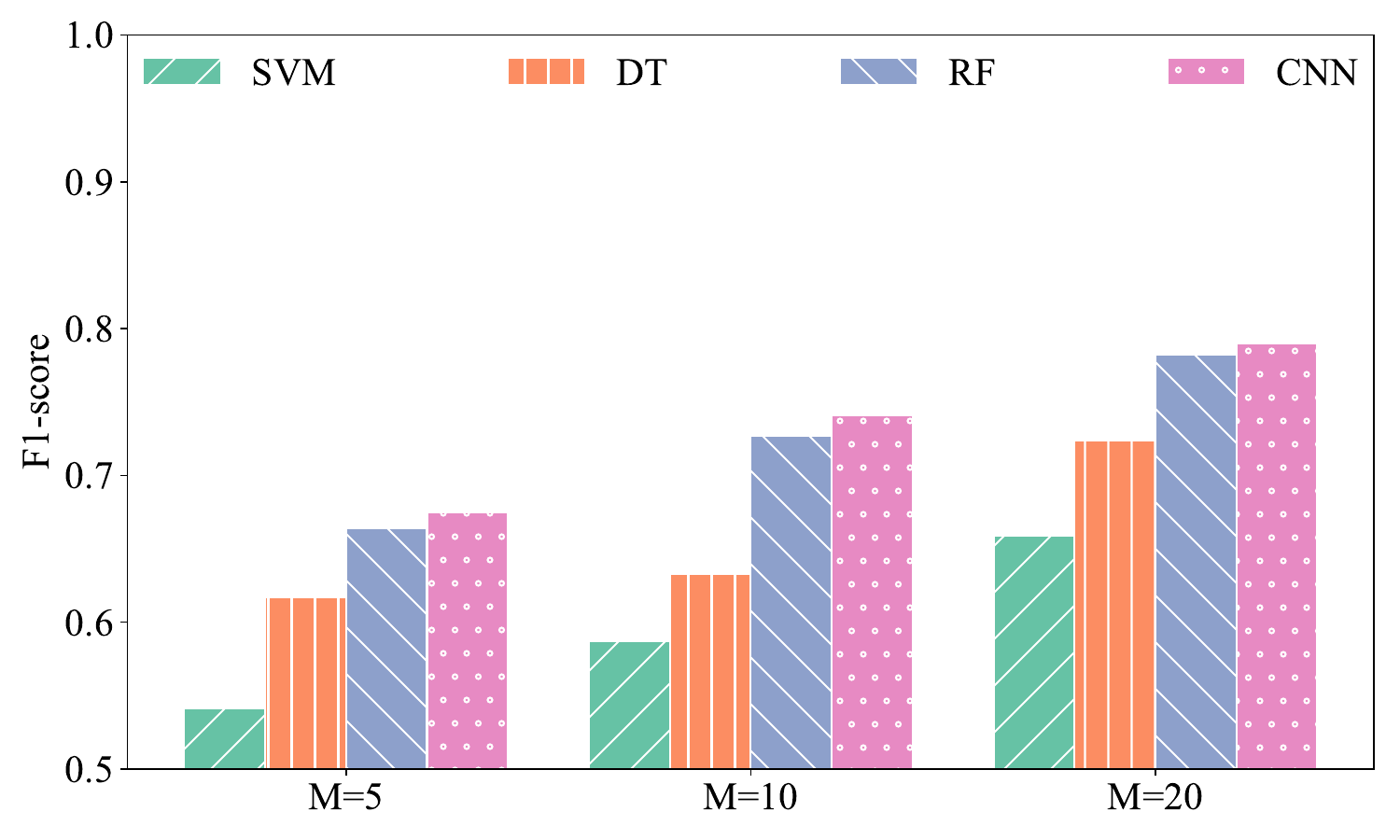}
		\label{2_1}
	}%
	\subfigure[]{
		\centering
		\includegraphics[width=0.32\linewidth]{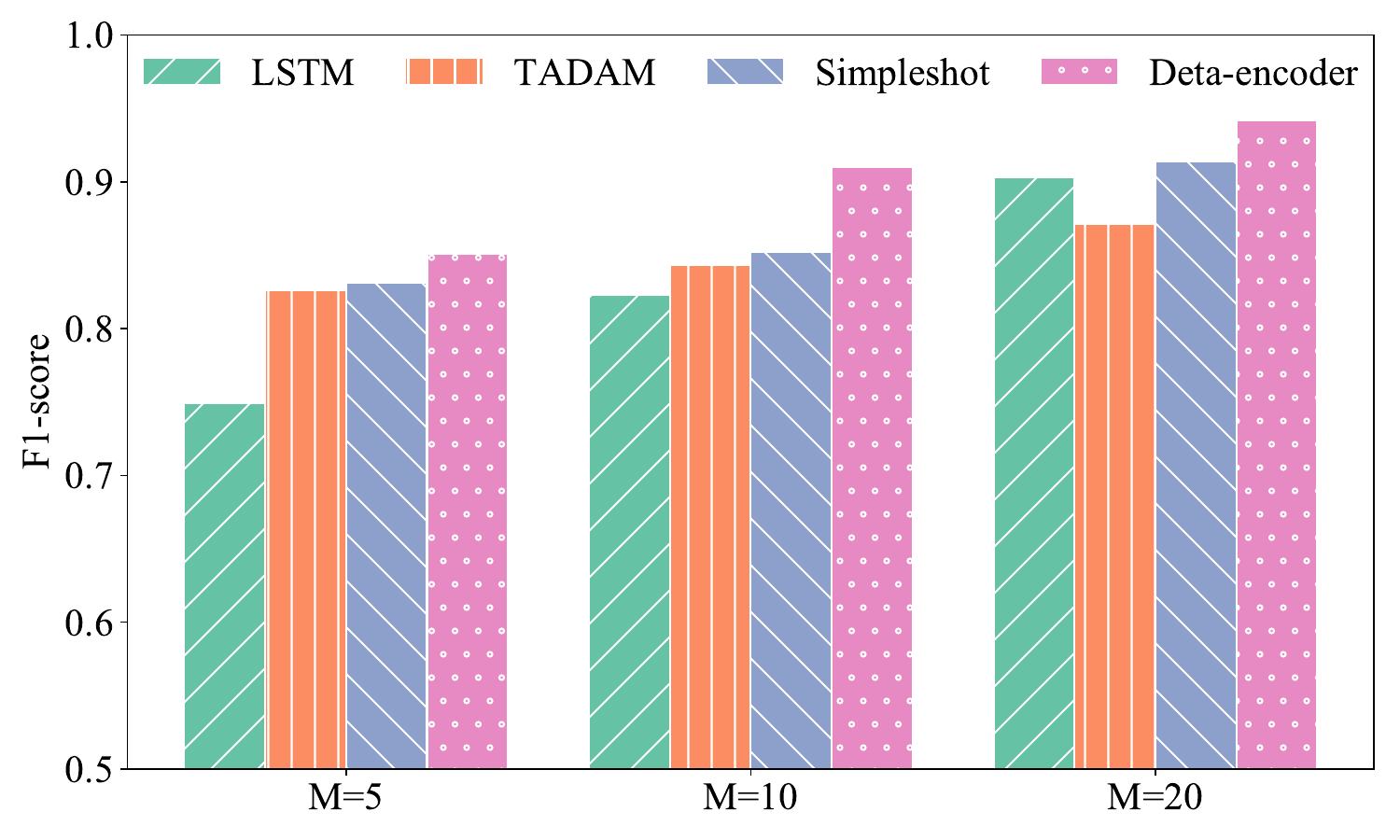}
		\label{2_2}
	}%
	\subfigure[]{
		\centering
		\includegraphics[width=0.32\linewidth]{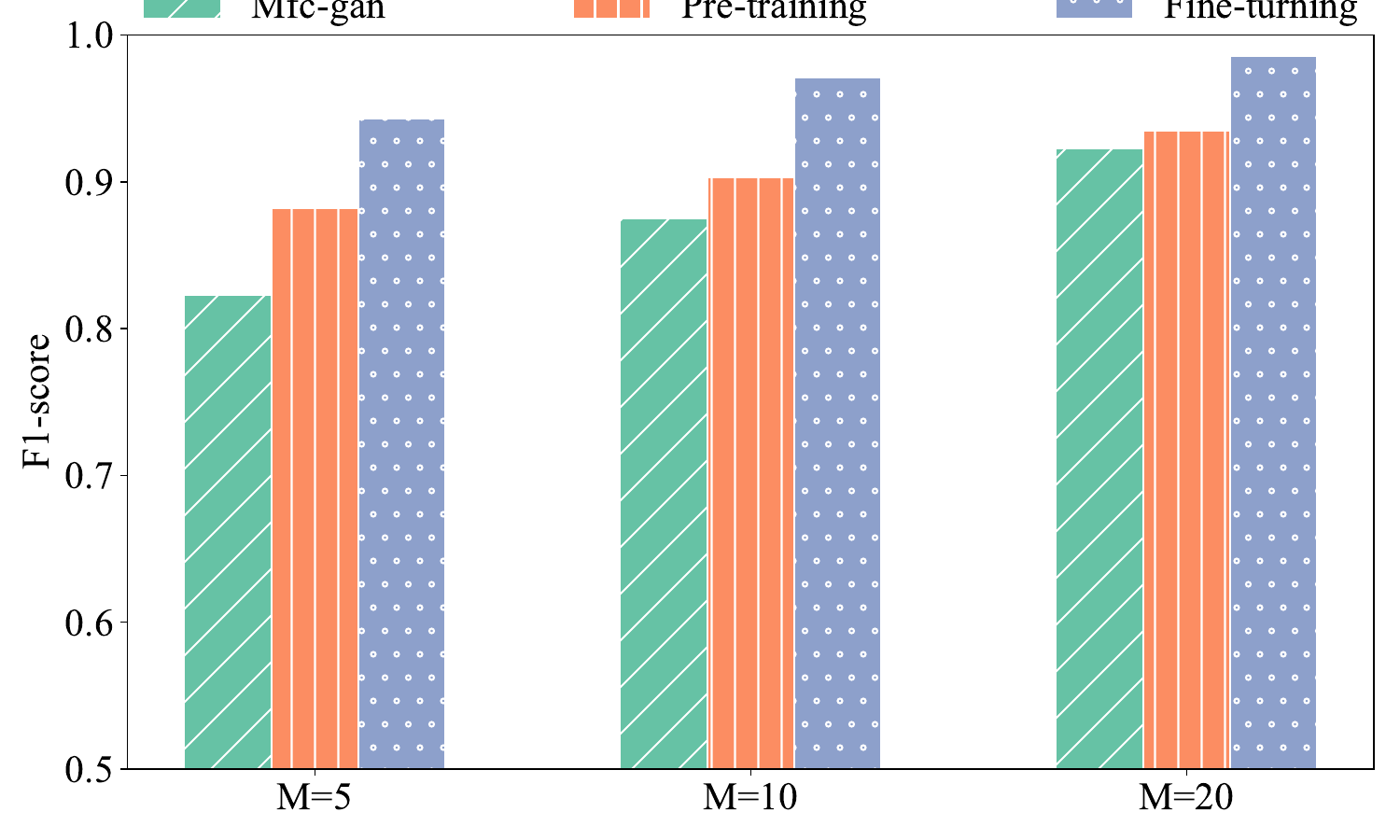}
		\label{2_3}
	}%
	\centering
	\caption{Compare Meta-UAD with existing anomaly detection models on the M-shot anomaly detection task. The results were collected on the CIC-AndMal2017 dataset.}
	\label{plot_2_1}
\end{figure*}

{\it Evaluation metrics.} Considering the data imbalanced condition in a multiclass classification problem, to efficiently judge our scheme's performance in detecting new anomaly classes with few labeled samples, we use accuracy and F1-score as the criterion (same as prior work \cite{MAML, MetaL2}). F1-score is a critical metric to verify whether data are correctly and efficiently classified. 

{\it Baselines.} Five standard classification models are selected as comparison algorithms: Support Vector Machine (SVM) \cite{SVM}, C4.5 decision tree (DT) \cite{DT}, Random Forest (RF) \cite{RF}, convolutional neural network (CNN) \cite{CNN}, and Long Shot-Term Memory (LSTM) \cite{LSTM}. Similar, two metric-based models and two generative-based models are selected as comparison algorithms: Simpleshot\cite{Simpleshot}, TADAM\cite{TADAM}, Delta-encoder\cite{Delta-encoder}, and Mfc-gan\cite{Mfcgan}. In the training phase, each algorithm uses all training datasets for parameter optimization. Moreover, we also define an additional baseline. {\it Pre-training:} the model learning from the meta-learning phase is directly applied to the meta-test phase without any adaptation. {\it Fine-tuning:} the model learning from the meta-learning phase is applied to the meta-test phase by a few iteration steps, which is the standard Meta-UAD.

\subsection{Meta-UAD on M-shot anomaly detection} 
In this subsection, to verify that Meta-UAD can accurately detect new anomaly classes with few labeled samples, we compare Meta-UAD with the existing models on the M-shot social multimedia traffic anomaly detection in terms of F1-score. In the CIC-AndMal2017 dataset, we randomly select 30 classes from 42 anomaly families as the training set $\mathcal{C}_{tr}$ and the remaining classes as the testing set $\mathcal{C}_{test}$. In each episode of meta-training phase, Meta-UAD samples $M$ ($M=5,10,20$) labeled flows per class $\mathcal{C}_{tr}^i$ to construct the support set $\mathcal{D}^{sup}_i$ and the validation set $\mathcal{D}^{val}_i$. We construct the task $\mathcal{T}_i = \{\mathcal{D}^{sup}_i,\mathcal{D}^{var}_i\}$ for each class $\mathcal{C}_{tr}^i$. When $M$ is equal to 5, 10, 20, respectively, we finally can obtain three meta-models through inner update iteration and outer update iteration. For existing models, during the training phase, we use the entire training dataset for model optimization until we get the optimal model on the considered evaluation metrics. During the testing phase, Meta-UAD first uses a pre-training meta-model to adapt to new anomaly classes $\mathcal{C}_{test}^i$ with $M$ labeled samples through few iterative steps. Then, we resample $M$ ($M=5,10,20$) labeled flows of class $\mathcal{C}_{test}^i$ in the testing set $\mathcal{C}_{test}$ for comparing between Meta-UAD and existing models. Since the sample number in the testing set is minimal, we choose to take the mean value of multiple experiments (100 times) to achieve the reliability of the results. The experimental results are shown in Fig. \ref{plot_2_1}.

\begin{table}[t]
	\centering
	\caption{Standard anomaly detection in CIC-AndMal2017.}
	\label{2}
	\begin{tabular}{c|ccc}
		\toprule
		Category & Model & Accuracy & F1\\ \midrule
		\multirow{5}{*}{\thead{Meachine learning\\based}} 
		& SVM & 0.683 & 0.696 \\
		& DT & 0.976 & 0.974 \\
		& RF & 0.979 & 0.981\\ 
		& CNN & 0.916& 0.912 \\
		& LSTM & 0.964 & 0.967 \\ 
		\multirow{2}{*}{\thead{Metric-based}} 
		& Simpleshot & 0.931 & 0.939 \\
		& TADAM & 0.933 & 0.936 \\
		\multirow{2}{*}{\thead{Generative-based}} 
		& Delta-encoder & 0.973& 0.968  \\
		& Mfc-gan& 0.913& 0.908 \\ \midrule
		Ours & Meta-UAD & {\bf 0.982} & {\bf 0.987}  \\
		\bottomrule
	\end{tabular}%
\end{table}

\begin{table}[t]
	\centering
	\caption{Standard anomaly detection in CIC-IDS2017.}
	\label{3}
	\begin{tabular}{c|ccc}
		\toprule
		Category & Model & Accuracy & F1\\ \midrule
		\multirow{5}{*}{\thead{Meachine learning\\based}} 
		& SVM &0.647 &0.731 \\
		& DT &0.954 &0.943 \\
		& RF &0.973 &0.969\\ 
		& CNN &0.895 &0.903 \\
		& LSTM  &0.953 &0.960 \\ 
		\multirow{2}{*}{\thead{Metric-based}} 
		& Simpleshot &0.927 &0.929 \\
		& TADAM &0.919 &0.928 \\
		\multirow{2}{*}{\thead{Generative-based}} 
		& Delta-encoder  &0.961 &0.957  \\
		& Mfc-gan &0.898 &0.904\\ \midrule
		Ours & Meta-UAD &{\bf 0.974} &{\bf 0.979}  \\
		\bottomrule
	\end{tabular}%
\end{table}
Our scheme is far superior to the machine-learning based models in terms of F1-score. Those models need to be trained in a large number of training samples, so to realize high-precision classification by summarizing the rules that distinguish anomaly traffic from normal traffic. They heavily depend on the dataset's distribution and only can achieve good results for a specific anomaly class set. Therefore, those models will perform extremely poorly for a new anomaly class. Since the anomaly class in the testing set has few labeled samples, those models can produce an inductive learning bias towards the majority classes. 

Our scheme is also better than metric-based and generative-based models in terms of F1-score. In the training phase, metric- and generative-based models learn a sample generator using only seen class data and do not explicitly learn to generate the new class samples. If we detect many new classes, the number of synthesizing samples would be unfeasibly estimated, and this operation will drastically increase computational complexity and energy consumption. In the testing phase, when the new anomaly class is significantly different from anomaly classes in the training set, the performance of those models is still poor.

We can find that the fine-tuning model is better than the pre-trained model. Before model testing, the fine-tuning model needs a few iteration steps to adjust parameters and adapt to the specific new anomaly class. This operation is one of the core advantages of Meta-UAD.

\begin{figure*}[t]
	\centering 
	\begin{minipage}{0.32\linewidth}
		\centering
		\includegraphics[width=0.9\linewidth]{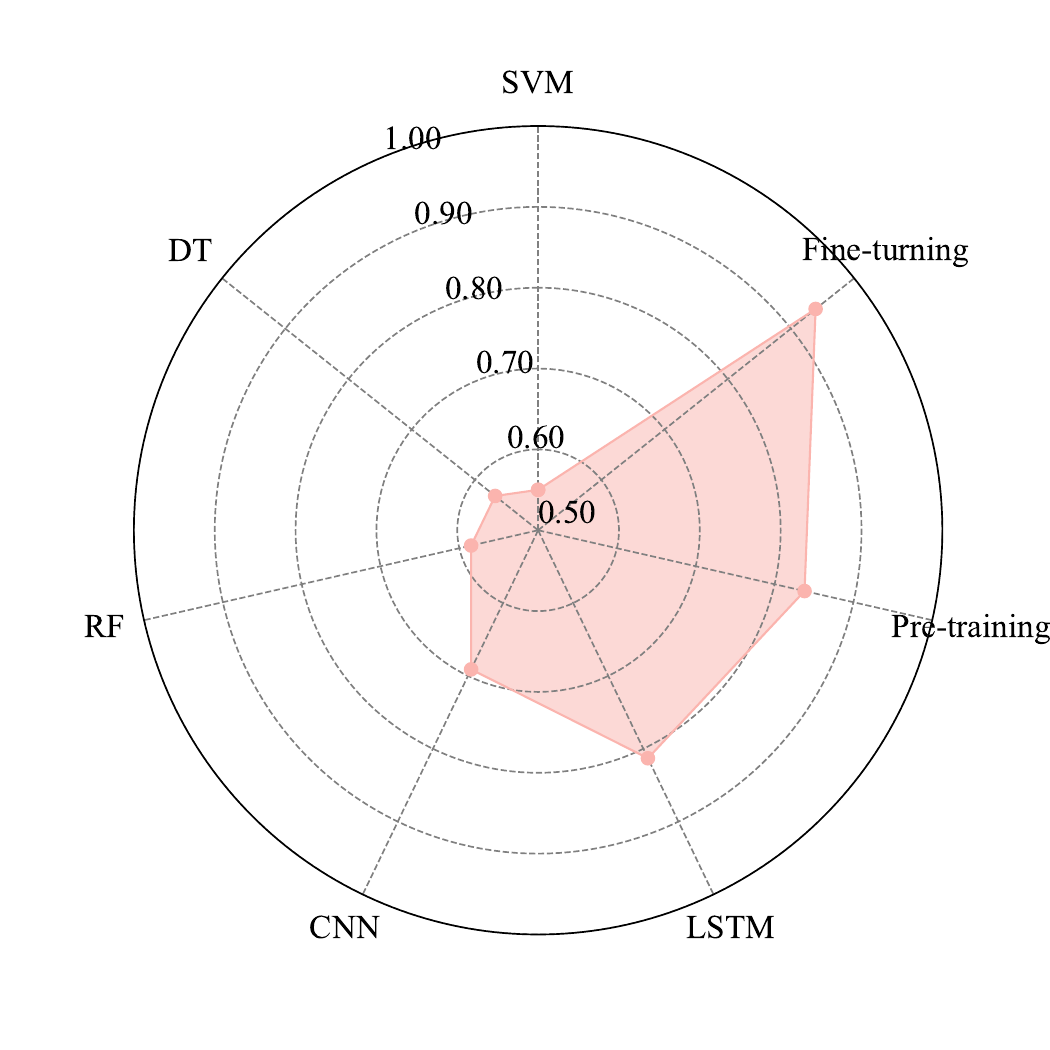}
		\caption{Compare our scheme with existing models in cross-dataset in terms of F1-score. The results were collected on the CIC-IDS2017 dataset.}
		\label{plot_3_1}
	\end{minipage}
	\begin{minipage}{0.32\linewidth}
		\centering
		\includegraphics[width=0.9\linewidth]{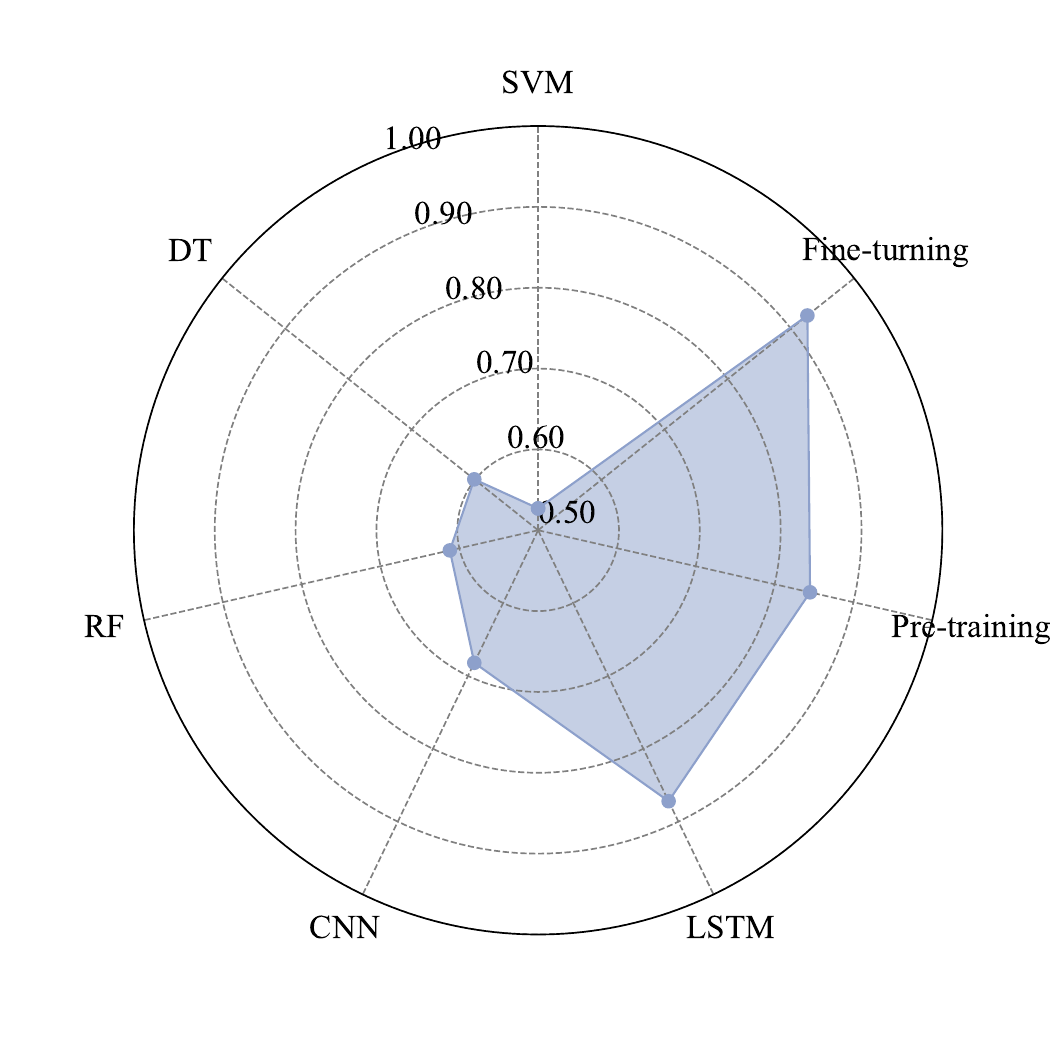}
		\caption{Compare our scheme with existing models in cross-dataset in terms of F1-score. The results were collected on the CIC-AndMal2017 dataset.}
		\label{plot_3_2}
	\end{minipage}
	\begin{minipage}{0.32\linewidth}
		\centering
		\includegraphics[width=0.9\linewidth]{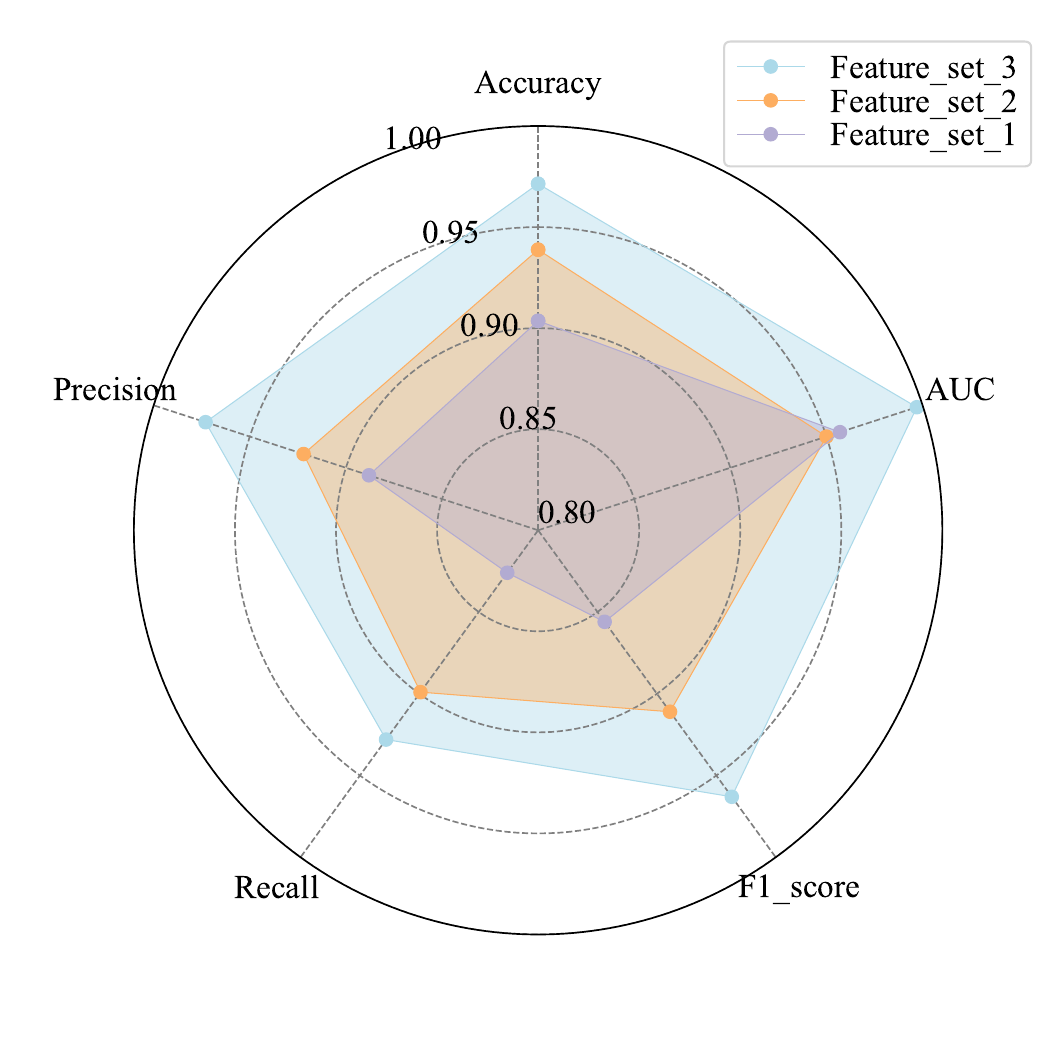}
		\caption{Compare the performence between three feature sets. The results were collected on the CIC-AndMal2017 dataset.}
		\label{plot_4}
	\end{minipage}
\end{figure*}

\subsection{Meta-UAD on standard anomaly detection} 
Meta-UAD's goal is to detect new anomaly classes with few labeled samples accurately. In real scenarios, existing well-known anomaly families still dominate the anomaly traffic. So we also verify that Meta-UAD can be applied to the standard anomaly detection problem: the training set and testing set contain many anomaly classes with large labeled samples and anomaly types in the testing set are the same as those in the training set. During the training phase, all models use the entire training dataset for model optimization until we get the optimal model on the considered evaluation metrics (accuracy and F1-score). For Meta-UAD, the sample number of support set/validation set equals 20 in each iteration. During the testing phase, meta-UAD use the pre-training meta-model for anomaly detection. We implemented 10-fold cross-validation in this experiment. The experimental results are collected on the CIC-AndMal2017 and the CIC-IDS2017 datasets.

From TABLE \ref{2} and TABLE \ref{3}, we can find that our scheme performs as well as others. In a large of training samples, the existing models can comprehensively summarize and learn how to distinguish anomaly traffic from normal traffic. So in the testing set with the same anomaly distribution, those models can achieve extremely high anomaly detection accuracy. Our scheme learns a meta-model through internal update iterations and external update iterations, which can accurately detect existing anomaly classes and generalize quickly to new anomaly classes. So our scheme also performs well in standard anomaly detection problems.

\subsection{Generalization} 
In this subsection, to verify that Meta-UAD has a strong generalization, we compared it with existing models in a cross-dataset in terms of F1-score. We select all anomaly families in the CIC-AndMal2017 dataset as the training set and anomaly families in the CIC-IDS2017 dataset as the testing set. In each episode of the meta-training phase, Meta-UAD samples 20 ($M=20$) labeled flows per class to construct the support set and the validation set. We can finally obtain one meta-model through inner update iteration and outer update iteration. For existing models, during the training phase, we use the entire training dataset for model optimization until we get the optimal model on the considered evaluation metrics. During the testing phase, Meta-UAD first uses the pre-training meta-model to adapt to new anomaly classes with 20 ($M=20$) labeled samples through few iterative steps. Then, we resample 20 ($M=20$) labeled flows per class in the testing set to compare Meta-UAD and existing models. Since the sample number in the testing set is minimal, we choose to take the mean value of multiple experiments (100 times) to achieve the reliability of the results. The experimental results are shown in Fig. \ref{plot_3_1}. On the contrary, the experimental results are shown in Fig. \ref{plot_3_2}.

We find that our scheme can still achieve optimal results in cross-dataset. Existing models heavily depend on the dataset's distribution and have poor generalization abilities to a new dataset. Our scheme needs to iterate some steps according to new anomaly samples in the meta-test phase, which can maximize the model's generalization and degrade the model's computational cost.

\subsection{Feature engineering} 
In this subsection, to verify the importance of feature engineering, we compared three feature sets: feature{\_}set{\_}1 includes 81 statistical features obtained through CICFlowMeter; feature{\_}set{\_}2 includes 33 statistical features obtained through cumulative importance ranking; feature{\_}set{\_}3 include 33 statistical features and five time-series features. By running these feature sets separately in Meta-UAD in terms of F1-score, the experimental results are shown in Fig. \ref{plot_4}.

We find that the performance of feature{\_}set{\_}2 is slightly higher than that of feature{\_}set{\_}1, but the performance of feature{\_}set{\_}3 is significantly higher than that of feature{\_}set{\_}1. This is because feature engineering can reduce the impact of indiscriminate features. Time-series features can provide a more comprehensive and fine-grained evaluation of the entire flow, which can obtain a more accurate flow description. Therefore, time-series relationships between packets are also critical for anomaly detection.

\section{Related Work}
This section reviews existing work on social multimedia traffic anomaly detection, few-shot learning, and meta-learning.

\subsection{Anomaly detection in social multimedia traffic} 
The goal of anomaly detection in user-level social multimedia traffic is to identify anomaly behaviors automatically by learning exclusively from normal traffic\cite{TNSM3,TNSM4,TNSM5,TNSM6}. In general, Existing social multimedia traffic anomaly detection models can be grouped into three categories based on the main techniques they rely on: machine learning (ML) based, metric-learning based, and generative-based models. 

Supervised ML models train a bi\setcounter{table}{0}nary or multiclass classifier by using labeled training datasets, which include Support Vector Machine (SVM)\cite{SVM}, Decision Tree (DT)\cite{DT}, Random Forest (RF)\cite{RF} and so on. Each model's input is a feature matrix in which rows and columns represent samples and feature values. The model's performance can be optimal in a balanced dataset by providing many relevant features. Unsupervised ML models group similar measurements in a single cluster and mark smaller clusters as anomaly classes, including K-means, Fuzzy C-means, Gaussian Mixture Models, etc. Deep ML models train a neural network to detect anomaly traffic by using back-propagation to calculate gradients and update nodes' weights, which include Deep Neural Network (DNN)\cite{DNN}, Convolutional Neural Network (CNN)\cite{CNN}, Long Shot-Term Memory (LSTM)\cite{LSTM} and so on. In contrast to supervised ML models, they do not build a feature matrix separately but identify important neurons through adjusting model weights in each iteration. They may improve classification precision by adjusting the number of hidden nodes and layers, activation functions, learning rate, or optimization algorithms. 

The latter two models reconstruct the current dataset and then use ML models for anomaly detection. The metric-learning based models can learn a non-linear metric to synthesize anomaly samples by distance to a class prototype or a class sub-space, such as Simpleshot\cite{Simpleshot} and TADAM\cite{TADAM}. The generative-based models can learn a sample generator only from seen class samples to synthesize new classes using Generative Adversarial Networks (GAN), such as Delta-encoder\cite{Delta-encoder} and Mfc-gan\cite{Mfcgan}. They contain an encoder and a decoder, which can generate new samples by modeling and learning the distribution of current datasets. Such as Delta-encoder based on a modified auto-encoder learns to synthesize new samples for a new category just by seeing a few examples from it. This model learns to map a novel sample instance to a concept and relates that concept to the existing ones in the concept space; using these relationships generates new instances by interpolating among the concepts to help learn.

\subsection{Few-Shot learning} 
Few-shot learning aims to learn representations that generalize well to unseen classes with only a few samples. Few-shot learning research can be divided into three categories: metric-learning based, generative-based models, and meta-learning based models. {\it Metric-learning based} models\cite{FSL1, FSL2, FSL3, FSL4, Simpleshot}, a non-linear metric is optimized based on seen classes and applied to unseen few-shot tasks. In \cite{FSL1}, using the distance to a class sub-space as the metric. As opposed to \cite{FSL1} and the model\cite{FSL2} try to optimize a sub-space for each class, which aims to seek a single sub-space optimally adapted to the few-shot task. Simpleshot \cite{Simpleshot} uses a nearest-neighbor classifier to achieve state-of-the-art results without pre-training. {\it Generative-based} models \cite{Delta-encoder, FSL7, FSL5, FSL8, FSL6, FSL9}, it generates more training examples for the few-shot learning task. Delta-encoder \cite{Delta-encoder} based on a modified auto-encoder learns to synthesize new samples for an unseen category just by seeing a few examples from it. In \cite{FSL5}, this model learns to construct an instance map and uses these relationships to generate new instances. However, those models do not discover a generic optimal parameter that can easily be generalized for seen and unseen class samples.

\subsection{Meta-learning}
Meta-learning has been an effective solution to the few-shot learning problem in practice. The meta-learning research can be divided into three classes: metric-based, model-based, and optimization-based models. {\it Metric-based} models \cite{ML-1,ML-2,ML-3,ML-4} learn a metric by reducing the intra-class variations while training on main categories. For example, Siamese \cite{ML-1} studies neural networks which can naturally rank similarities between inputs. MMN\cite{ML-2} augment neural networks with external memories to study a metric. RN\cite{ML-3} learns a distance metric between small sample classes in training, then can detect unseen classes by computing distance without further updating the network. {\it Model-based} approaches \cite{ML-5, ML-6, ML-9} can achieve rapid parameter updating of the model during training steps. Lastly, {\it optimization-based} models \cite{ML-7, ML-8, ML-10} modify the optimization algorithm for quick adaptation. These models can quickly adapt to a new task through the meta-update scheme among multiple tasks during parameter optimization. In social multimedia traffic anomaly detection, we follow a similar optimization-based meta-learning model Meta-SGD \cite{Meta-SGD} and apply it to the much more challenging anomaly detection task with new classes.

\section{CONCLUSION}
Accuracy anomaly detection in user-level social multimedia traffic is crucial for privacy security. Anomaly detection aims to automatically identify anomaly behaviors by learning exclusively from normal traffic. Compared with existing models that passively detect specific anomaly classes with large labeled training samples, user-level social multimedia traffic require real-time detection, analysis, and interception of all anomalies. They contain sizeable new anomaly classes with few labeled samples and have an imbalance, self-similar, and data-hungry nature. Many existing models have been proposed to address this problem. However, most existing models are usually data-hungry and have limited generalization abilities to new anomaly classes. So we proposed Meta-UAD, a meta-learning scheme for user-level social multimedia traffic anomaly detection. This scheme relies on the episodic training paradigm and learns from the collection of K-way-M-shot classification tasks, which can use the pre-trained model to adapt new anomaly classes with few samples going through few iteration steps. Since the social multimedia traffic at the user level emerges from a complex interaction process of users and application protocol, we further developed a feature extractor to improve scheme performance. By experiment evaluation, our scheme is better than others. In the future, we hope to deploy our scheme into a real gateway.

\bibliographystyle{./IEEEtran}
\bibliography{./IEEEabrv, ./TNSE}

% Generated by IEEEtran.bst, version: 1.12 (2007/01/11)
\begin{thebibliography}{10}
\providecommand{\url}[1]{#1}
\csname url@samestyle\endcsname
\providecommand{\newblock}{\relax}
\providecommand{\bibinfo}[2]{#2}
\providecommand{\BIBentrySTDinterwordspacing}{\spaceskip=0pt\relax}
\providecommand{\BIBentryALTinterwordstretchfactor}{4}
\providecommand{\BIBentryALTinterwordspacing}{\spaceskip=\fontdimen2\font plus
\BIBentryALTinterwordstretchfactor\fontdimen3\font minus
  \fontdimen4\font\relax}
\providecommand{\BIBforeignlanguage}[2]{{%
\expandafter\ifx\csname l@#1\endcsname\relax
\typeout{** WARNING: IEEEtran.bst: No hyphenation pattern has been}%
\typeout{** loaded for the language `#1'. Using the pattern for}%
\typeout{** the default language instead.}%
\else
\language=\csname l@#1\endcsname
\fi
#2}}
\providecommand{\BIBdecl}{\relax}
\BIBdecl

\bibitem{base3}
{Shi, Lumin and Li, Jun and Zhang, Mingwei and Reiher, Peter}, ``{On Capturing
  DDoS Traffic Footprints on the Internet},'' \emph{{IEEE Transactions on
  Dependable and Secure Computing}}, vol.~19, no.~4, pp. 2755--2770, 2022.

\bibitem{TNSM1}
A.~Putina and D.~Rossi, ``Online anomaly detection leveraging stream-based
  clustering and real-time telemetry,'' \emph{IEEE Transactions on Network and
  Service Management}, vol.~18, no.~1, pp. 839--854, 2021.

\bibitem{TNSM2}
A.~Dridi, C.~Boucetta, S.~E. Hammami, H.~Afifi, and H.~Moungla, ``Stad:
  Spatio-temporal anomaly detection mechanism for mobile network management,''
  \emph{IEEE Transactions on Network and Service Management}, vol.~18, no.~1,
  pp. 894--906, 2021.

\bibitem{base4}
{Xiao, Xi and Xiao, Wentao and Li, Rui and Luo, Xiapu and Zheng, Haitao and
  Xia, Shutao}, ``{EBSNN: Extended Byte Segment Neural Network for Network
  Traffic Classification},'' \emph{{IEEE Transactions on Dependable and Secure
  Computing}}, vol.~19, no.~5, pp. 3521--3538, 2022.

\bibitem{TNSM7}
L.~Nie, L.~Zhao, and K.~Li, ``Robust anomaly detection using reconstructive
  adversarial network,'' \emph{IEEE Transactions on Network and Service
  Management}, vol.~18, no.~2, pp. 1899--1912, 2021.

\bibitem{SVM}
{Suykens, Johan AK and Vandewalle, Joos}, ``{Least Squares Support Vector
  Machine Classifiers},'' \emph{{Neural Processing Letters}}, vol.~9, no.~3,
  pp. 293--300, 1999.

\bibitem{DT}
{Wang, Shanshan and Chen, Zhenxiang and Zhang, Lei and Yan, Qiben and Yang, Bo
  and Peng, Lizhi and Jia, Zhongtian}, ``{TrafficAV: An Effective and
  Explainable Detection of Mobile Malware Behavior using Network Traffic},'' in
  \emph{{IWQoS}}, 2016, pp. 1--6.

\bibitem{RF}
{Liaw, Andy and Wiener, Matthew}, ``{Classification and Regression by Random
  Forest},'' \emph{{R News}}, vol.~2, no.~3, pp. 18--22, 2002.

\bibitem{CNN}
{Min, Erxue and Long, Jun and Liu, Qiang and Cui, Jianjing and Chen, Wei},
  ``{TR-IDS: Anomaly-based Intrusion Detection through Text-convolutional
  Neural Network and Random Forest},'' \emph{{Security and Communication
  Networks}}, vol. 2018, pp. 1--9, 2018.

\bibitem{LSTM}
{Lotfollahi, Mohammad and Siavoshani, Mahdi Jafari and Zade, Ramin Shirali
  Hossein and Saberian, Mohammdsadegh}, ``{Deep Packet: A Novel Approach for
  Encrypted Traffic Classification Using Deep Learning},'' \emph{{Soft
  Computing}}, vol.~24, no.~3, pp. 1999--2012, 2020.

\bibitem{Simpleshot}
{Wang, Yan and Chao, Wei-Lun and Weinberger, Kilian Q and van der Maaten,
  Laurens}, ``{Simpleshot: Revisiting Nearest-neighbor Classification for
  Few-shot Learning},'' \emph{{ArXiv Preprint}}, 2019.

\bibitem{TADAM}
{Oreshkin, Boris and Rodriguez, Pau and Lacoste, Alexandre}, ``{TADAM: Task
  Dependent Adaptive Metric for Improved Few-shot Learning},'' in
  \emph{{NIPS}}, 2018, p. 7352.

\bibitem{Delta-encoder}
{Schwartz, Eli and Karlinsky, Leonid and Shtok, Joseph and Harary, Sivan and
  Marder, Mattias and Feris, Rogerio and Kumar, Abhishek and Giryes, Raja and
  Bronstein, Alex M}, ``{Delta-encoder: An Effective Sample Synthesis Method
  for Few-shot Object Recognition},'' in \emph{{NIPS}}, 2018.

\bibitem{Mfcgan}
{Ali-Gombe, Adamu and Elyan, Eyad}, ``{Mfc-gan: Class-imbalanced Dataset
  Classification Using Multiple Fake Class Generative Adversarial Network},''
  \emph{{Neurocomputing}}, vol. 361, pp. 212--221, 2019.

\bibitem{FSL1}
{Hersche, Michael and Karunaratne, Geethan and Cherubini, Giovanni and Benini,
  Luca and Sebastian, Abu and Rahimi, Abbas}, ``{Constrained Few-shot
  Class-incremental Learning},'' in \emph{{CVPR}}, 2022, pp. 9047--9057.

\bibitem{FSL2}
{Zhang, Bo and Ye, Hancheng and Yu, Gang and Wang, Bin and Wu, Yike and Fan,
  Jiayuan and Chen, Tao}, ``{Sample-Centric Feature Generation for
  Semi-Supervised Few-Shot Learning},'' \emph{{IEEE Transactions on Image
  Processing}}, vol.~31, pp. 2309--2320, 2022.

\bibitem{FSL3}
{Li, Wei-Hong and Liu, Xialei and Bilen, Hakan}, ``{Cross-domain Few-shot
  Learning with Task-specific Adapters},'' in \emph{{CVPR}}, 2022, pp.
  7151--7160.

\bibitem{FSL4}
{Phaphuangwittayakul, Aniwat and Guo, Yi and Ying, Fangli}, ``{Fast Adaptive
  Meta-Learning for Few-Shot Image Generation},'' \emph{{IEEE Transactions on
  Multimedia}}, vol.~24, pp. 2205--2217, 2022.

\bibitem{base1}
{Lu, Yiwei and Yu, Frank and Reddy, Mahesh Kumar Krishna and Wang, Yang},
  ``{Few-shot scene-adaptive anomaly detection},'' in \emph{{ECCV}}, 2020, pp.
  125--141.

\bibitem{MAML}
{Finn, Chelsea and Abbeel, Pieter and Levine, Sergey}, ``{Model-Agnostic
  Meta-Learning for Fast Adaptation of Deep Networks},'' in \emph{{ICML}},
  2017, pp. 1126--1135.

\bibitem{Meta-SGD}
{Li, Zhenguo and Zhou, Fengwei and Chen, Fei and Li, Hang}, ``{Meta-SGD:
  Learning to Learn Quickly for Few-shot Learning},'' \emph{{ArXiv Preprint}},
  2017.

\bibitem{AutoEncoder1}
{Min, Erxue and Long, Jun and Liu, Qiang and Cui, Jianjing and Cai, Zhiping and
  Ma, Junbo}, ``{SU-IDS: A Semi-supervised and Unsupervised Framework for
  Network Intrusion Detection},'' in \emph{{International Conference on Cloud
  Computing and Security}}, 2018, pp. 322--334.

\bibitem{CICFlowMeter}
{Lashkari, Arash Habibi and Draper-Gil, Gerard and Mamun, Mohammad Saiful Islam
  and Ghorbani, Ali A}, ``{Characterization of Tor Traffic Using Time based
  Features},'' in \emph{{ICISSP}}, 2017, pp. 253--262.

\bibitem{DNN}
{Wang, Xiaofei and Han, Yiwen and Leung, Victor CM and Niyato, Dusit and Yan,
  Xueqiang and Chen, Xu}, ``{Convergence of Edge Computing and Deep Learning: A
  Comprehensive Survey},'' \emph{{IEEE Communications Surveys and Tutorials}},
  vol.~22, no.~2, pp. 869--904, 2020.

\bibitem{CICAndMal2017}
{A. H. Lashkari and A. F. A. Kadir and L. Taheri and A. A. Ghorbani}, ``{Toward
  Developing a Systematic Approach to Generate Benchmark Android Malware
  Datasets and Classification},'' in \emph{{ICCST}}, 2018, pp. 1--7.

\bibitem{CICIDS2017}
{Iman Sharafaldin and Arash Habibi Lashkari and A. Ghorbani}, ``{Toward
  Generating a New Intrusion Detection Dataset and Intrusion Traffic
  Characterization},'' in \emph{{ICISSP}}, 2018.

\bibitem{TNSM3}
M.~Lyu, H.~H. Gharakheili, C.~Russell, and V.~Sivaraman, ``Hierarchical
  anomaly-based detection of distributed dns attacks on enterprise networks,''
  \emph{IEEE Transactions on Network and Service Management}, vol.~18, no.~1,
  pp. 1031--1048, 2021.

\bibitem{TNSM4}
P.~Satam and S.~Hariri, ``Wids: An anomaly based intrusion detection system for
  wi-fi (ieee 802.11) protocol,'' \emph{IEEE Transactions on Network and
  Service Management}, vol.~18, no.~1, pp. 1077--1091, 2021.

\bibitem{TNSM5}
I.~Siniosoglou, P.~Radoglou-Grammatikis, G.~Efstathopoulos, P.~Fouliras, and
  P.~Sarigiannidis, ``A unified deep learning anomaly detection and
  classification approach for smart grid environments,'' \emph{IEEE
  Transactions on Network and Service Management}, vol.~18, no.~2, pp.
  1137--1151, 2021.

\bibitem{TNSM6}
D.~C. Le and N.~Zincir-Heywood, ``Anomaly detection for insider threats using
  unsupervised ensembles,'' \emph{IEEE Transactions on Network and Service
  Management}, vol.~18, no.~2, pp. 1152--1164, 2021.

\bibitem{FSL7}
{Zhuo, Linhai and Fu, Yuqian and Chen, Jingjing and Cao, Yixin and Jiang,
  Yu-Gang}, ``{TGDM: Target Guided Dynamic Mixup for Cross-Domain Few-Shot
  Learning},'' in \emph{{ACM MM}}, 2022, p. 6368–6376.

\bibitem{FSL5}
{Wang, Haoxiang and Wang, Yite and Sun, Ruoyu and Li, Bo}, ``{Global
  Convergence of MAML and Theory-Inspired Neural Architecture Search for
  Few-Shot Learning},'' in \emph{{CVPR}}, 2022, pp. 9787--9798.

\bibitem{FSL8}
{Zhang, Xingxing and Liu, Zhizhe and Yang, Weikai and Wang, Liyuan and Zhu,
  Jun}, ``{The More, The Better? Active Silencing of Non-Positive Transfer for
  Efficient Multi-Domain Few-Shot Classification},'' in \emph{{ACM MM}}, 2022,
  p. 1993–2001.

\bibitem{FSL6}
{Huang, Huaxi and Zhang, Junjie and Zhang, Jian and Xu, Jingsong and Wu,
  Qiang}, ``{Low-Rank Pairwise Alignment Bilinear Network For Few-Shot
  Fine-Grained Image Classification},'' \emph{IEEE Transactions on Multimedia},
  vol.~23, pp. 1666--1680, 2021.

\bibitem{FSL9}
{Wang, Shuo and Zhang, Xinyu and Hao, Yanbin and Wang, Chengbing and He,
  Xiangnan}, ``{Multi-Directional Knowledge Transfer for Few-Shot Learning},''
  in \emph{{ACM MM}}, 2022, p. 3993–4002.

\bibitem{ML-1}
{Cai, Qi and Pan, Yingwei and Yao, Ting and Yan, Chenggang and Mei, Tao},
  ``{Memory Matching Networks for One-Shot Image Recognition},'' in
  \emph{{CVPR}}, 2018, pp. 4080--4088.

\bibitem{ML-2}
{Vinyals, Oriol and Blundell, Charles and Lillicrap, Timothy and koray
  kavukcuoglu and Wierstra, Daan}, ``{Matching Networks for One Shot
  Learning},'' in \emph{{NIPS}}, 2016, pp. 3630--3638.

\bibitem{ML-3}
{Sung, Flood and Yang, Yongxin and Zhang, Li and Xiang, Tao and Torr, Philip
  and Hospedales, Timothy}, ``{Learning to Compare: Relation Network for
  Few-Shot Learning},'' in \emph{{CVPR}}, 2018, pp. 1199--1208.

\bibitem{ML-4}
{Baik, Sungyong and Choi, Janghoon and Kim, Heewon and Cho, Dohee and Min,
  Jaesik and Lee, Kyoung Mu}, ``{Meta-Learning with Task-Adaptive Loss Function
  for Few-Shot Learning},'' in \emph{{ICCV}}, 2021, pp. 9445--9454.

\bibitem{ML-5}
{Zhang, Pei and Bai, Yunpeng and Wang, Dong and Bai, Bendu and Li, Ying}, ``{A
  Meta-Learning Framework for Few-Shot Classification of Remote Sensing
  Scene},'' in \emph{{ICASSP}}, 2021, pp. 4590--4594.

\bibitem{ML-6}
{Lu, Changsheng and Koniusz, Piotr}, ``{Few-shot Keypoint Detection with
  Uncertainty Learning for Unseen Species},'' in \emph{{CVPR}}, 2022, pp.
  19\,394--19\,404.

\bibitem{ML-9}
{Guan, Jiechao and Lu, Zhiwu}, ``{Fast-Rate PAC-Bayesian Generalization Bounds
  for Meta-Learning},'' in \emph{{ICML}}, 2022, pp. 7930--7948.

\bibitem{ML-7}
{Zhang, Huaiwen and Qian, Shengsheng and Fang, Quan and Xu, Changsheng},
  ``{Multi-Modal Meta Multi-Task Learning for Social Media Rumor Detection},''
  \emph{{IEEE Transactions on Multimedia}}, vol.~24, pp. 1449--1459, 2022.

\bibitem{ML-8}
{Abbas, Momin and Xiao, Quan and Chen, Lisha and Chen, Pin-Yu and Chen,
  Tianyi}, ``{Sharp-MAML: Sharpness-Aware Model-Agnostic Meta Learning},'' in
  \emph{{ICML}}, 2022, pp. 10--32.

\bibitem{ML-10}
{Pong, Vitchyr H and Nair, Ashvin V and Smith, Laura M and Huang, Catherine and
  Levine, Sergey}, ``{Offline Meta-Reinforcement Learning with Online
  Self-Supervision},'' in \emph{{ICML}}, 2022, pp. 17\,811--17\,829.

\end{thebibliography}

\end{document}